\documentclass[conference,twocolumn,10pt]{IEEEtran}

\usepackage[utf8]{inputenc} %
\usepackage[T1]{fontenc}    %
\usepackage{url}            %
\usepackage{booktabs}       %
\usepackage{amsfonts}       %
\usepackage{nicefrac}       %
\usepackage{microtype}      %
\usepackage{amsmath}
\usepackage{graphicx}
\usepackage{subcaption}
\usepackage{multirow}
\usepackage{array}
\usepackage{wrapfig}
\usepackage{diagbox}
\usepackage[font=small,skip=5pt]{caption}

\DeclareMathOperator{\argmin}{arg\,min}

\usepackage{xcolor}

\newcommand{\calF}{\mathcal{F}}
\newcommand{\calA}{\mathcal{A}}
\newcommand{\calR}{\mathcal{R}}

\newcommand{\RR}{\mathbb{R}}
\newcommand{\tx}{\tilde{x}}

\newcommand{\para}[1]{{\vspace{2pt} \bf \noindent #1}}   

\newenvironment{packed_itemize}{
\begin{list}{\labelitemi}{\leftmargin=1em}
\setlength{\itemsep}{1pt}                                                           
\setlength{\parskip}{0pt}                                                                                 \setlength{\parsep}{0pt}                                                                                  \setlength{\headsep}{0pt}                                                                                 \setlength{\topskip}{0pt}                                                                                 \setlength{\topmargin}{0pt}                                                                               \setlength{\topsep}{0pt}                                                                                  \setlength{\partopsep}{0pt}                                                                               }{\end{list}}

\newtheorem{definition}{Definition}

\title{Assessing Privacy Risks from Feature Vector Reconstruction Attacks}
\author{
    \IEEEauthorblockN{
    Emily Wenger\IEEEauthorrefmark{1},
    Francesca Falzon\IEEEauthorrefmark{1}\IEEEauthorrefmark{2}, 
    Josephine Passananti\IEEEauthorrefmark{1}, 
    Haitao Zheng\IEEEauthorrefmark{1}, 
    Ben Y. Zhao\IEEEauthorrefmark{1}}
    \IEEEauthorblockA{\IEEEauthorrefmark{1} University of Chicago, \IEEEauthorrefmark{2} Brown University
    \\\{ewenger, josephinep, htzheng, ravenben\}@uchicago.edu, \{francesca\_falzon\}@brown.edu}
}
\begin{document}

\maketitle

\begin{abstract}
  In deep neural networks for facial recognition, feature vectors are
  numerical representations that capture the unique features of a given
  face. While it is known that a version of the original face can be
  recovered via ``feature reconstruction,'' we lack an understanding of the
  end-to-end privacy risks produced by these attacks. In this work, we
  address this shortcoming by developing metrics that meaningfully capture
  the threat of reconstructed face images. Using end-to-end experiments and
  user studies, we show that reconstructed face images enable
  re-identification by both commercial facial recognition systems and humans,
  at a rate that is at worst, a factor of four times higher than randomized
  baselines. Our results confirm that feature vectors should be recognized as
  Personal Identifiable Information (PII) in order to protect user privacy.
  
\end{abstract}

\section{Introduction}

Feature vectors allow quick identification of similar images, and both facial recognition
engines and anti-facial recognition tools rely on them. Given an image $x$, a well-trained deep neural network (DNN) $\calF$
maps $x$ to a feature vector $\calF(x) = v$ that represents $x$'s visual and/or semantic
features. Facial recognition systems use databases of feature vectors to produce identity matches via
vector comparisons~\cite{azure, amazon_rek, he2016deep}. Anti-facial recognition tools
like Fawkes, designed to prevent unwanted facial recognition, corrupt feature vectors to prevent identification~\cite{fawkes}. Some tools
also propose ``collaborative'' anti-facial recognition schemes, where
individuals share their feature vectors with others for enhanced
protection~\cite{foggy}.

The privacy risks of publicly sharing face feature vectors
are explicitly or implicitly assumed to be minimal. For example,
several facial recognition companies state on their websites
that, since face feature vectors are numerical, they will not leak
sensitive personal information even if the feature vector database is hacked~\cite{facefirst,anyvision}. Similarly, anti-facial recognition
tools with feature vector sharing schemes do not consider whether this sharing could compromise privacy. Such assumptions are concerning, especially given the significant
regulatory standards for protection of other biometric
data~\cite{GDPR, bipa}. Face feature vectors also contain biometric data, so
the privacy risks of using or sharing them must be better understood.

One obvious threat to face feature vector privacy comes from so-called ``feature
vector reconstruction'' (FVR) techniques. FVR transforms the feature vector $v$ back into the image $x$~\cite{Vec2Face, CoRR:ZhoSan16,
  CVPR:MahVed15, nbnet, blackbox20}. 
 Prior work on inverse biometrics has concluded that
reconstruction alone is a severe attack~\cite{gomez2020reversing,
  cao2014learning}. However, in the context of facial
recognition, FVR could lead to a more serious privacy threat -- deanonymization. Since the
goal of FVR is to recreate the original image $x$, reconstructed face images could be fed back into a facial
  recognition system and re-identified. In addition to the privacy
  loss from deanonymization, reidentified face images could be used
  to conduct further attacks, like creating Deepfakes or crafting a 3D model of the victim's
face to fool facial recognition systems~\cite{deepfake, 3dmask_paper, 3dmask_attack}.

FVR-enabled deanonymization poses a plausible and potentially significant privacy
risk, but assessing this threat requires a carefully designed metric. %
Prior work on face FVR measures the visual similarity between original
and reconstructed images, implicitly making a strong assumption that the attacker already knows the target's identity. In practice, the target identity
is unknown, and finding a target match visually would require inspecting a
huge set of images. To scale the attack, an attacker needs to use a
facial recognition engine to identify a small set of potential matches
for visual inspection. Because facial recognition engines map images to feature
vectors, the reconstructed images must have similar features to the target (to
be among the potential matches) and be visually similar enough so the attacker can identify
them from the matches. Thus, to measure the real-world privacy risk of
FVR-enabled deanonymization we believe we need to consider both feature space and visual similarity. %

To fill this gap, our work proposes two metrics. The first, {\em top-K
  matching accuracy}, assesses whether a reconstructed image would
lead a facial recognition engine to produce a set of matches
containing the true target identity. The second compares the {\em visual
matching accuracy} between reconstructed and true images via a user
study. Armed with these metrics, we explore the limits of FVR-enabled
deanonymization attacks using state-of-the-art FVR methods, real-world
facial recognition systems, and a user study. Despite the
feature-space noise that current FVR methods add to reconstructed
images, our findings show that FVR-enabled deanonymization poses
a near-term threat and that the research community must take the necessary steps to mitigate this threat.

{\bf Our Contributions.} We make the following contributions to the community's understanding of FVR-enabled deanonymization attacks.
\begin{packed_itemize}
\item We develop two concrete metrics -- topK matching and visual
  matching -- that assess the real-world privacy risk
  of FVR-enabled deanonymization attacks (\S\ref{sec:metrics}).
\item We evaluate the deanonymization success of four state-of-the-art FVR methods using
  commercial facial recognition systems and an IRB-approved user study
  (\S\ref{sec:feature_results}, \S\ref{sec:visual_results}) and
    find that deanonymization is possible.

\end{packed_itemize}

\begin{figure*}
  \centering
  \includegraphics[width=0.99\textwidth]{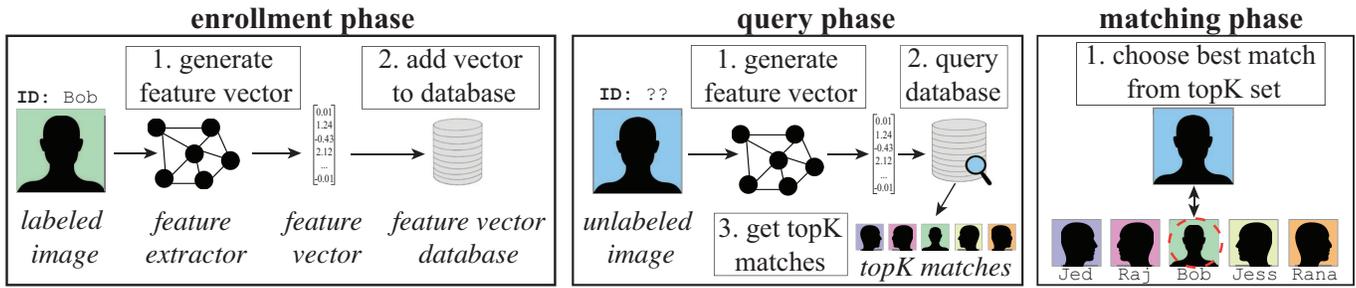}
  \caption{{\em Overview of three stages of facial recognition:
      enrollment, query, and matching.}%
  \label{fig:face_rec_overview}}
\end{figure*}

\section{Background}
\label{sec:back}

Before discussing the threat model and evaluation methodology of our
work, we first provide background information on feature vector
reconstruction and facial recognition. 

\para{Feature Vector Reconstruction.} FVR
methods use a reconstructor $\calR$ to transform a feature vector $\calF(x)
= v$ into a reconstructed image $\tx$. More formally, let $W,H,C,$ and $M$ be integers and let $\calF: \RR^{W\times H\times
  C}\to \RR^M$ be a function defined by a feature extractor that takes
as input an image $x$ and outputs a face feature vector $v$;
$v$ is an $M$-dimensional vector which represents $x$ in the feature
space of $\calF$. Given a feature vector $v$ we want to generate an image $\tx$, such that
$\calF(\tx)=v$. %
We thus define the problem of feature vector reconstruction as follows:

\noindent\rule{\linewidth}{0.4pt}
\begin{definition} Given black-box access to a model $\calF$ and a
  feature vector $v=\calF(x)$ of an unknown image $x$, the goal of
  \emph{feature vector reconstruction} (FVR) is to recover an
  image $\tx^*=\argmin_{\tx} \mathcal{L}(\calF(\tx),v)$, where
  $\mathcal{L}$ is some loss function. 
\end{definition}
\noindent\rule{\linewidth}{0.4pt}

The feature vector reconstructor $\calR_{\calF}$ is designed to invert feature vectors
produced by $\calF$, so $\calR_{\calF}(v) = \tx$. In practice, $\calR_{\calF}$ is
typically either a trained model (parametric method) or an
optimization procedure (nonparametric method).
{\em Parametric reconstruction methods} rely on a reconstruction model
$\calR$ trained specifically to invert vectors produced by a model
$\calF$~\cite{nbnet, Vec2Face, deeppoisoning21}. 
{\em Nonparametric reconstruction methods} use an iterative optimization
process $\calR$ to reconstruct images from $\calF$'s feature
vectors~\cite{blackbox20,eigenfaces}.

\para{Facial Recognition.} Since we evaluate FVR in the context of
facial recognition, we now briefly describe how modern facial
recognition systems work. Facial recognition works in three phases: {\em enrollment}, {\em query}, and {\em
 matching}, shown in Fig.~\ref{fig:face_rec_overview}. In the enrollment phase, labeled face images are transformed
into feature vectors and stored in a reference database. In the query phase, a user runs an unlabeled
image against the reference database, and the system returns the top-$K$ matches for the image, based on the similarity of their feature
vectors. Finally, in the matching phase, the user chooses the best match
from the top-$K$ set~\cite{deng2019arcface, he2016deep}. This general
setup is commonly used in commercial facial recognition systems~\cite{gao_2020}.

\para{Relationship between FVR and Facial Recognition.} The three
phases of facial recognition correspond to stages of FVR-enabled
deanonymization attacks. Attackers must first obtain feature vectors to
reconstruct. These could be obtained from leaked facial recognition reference
databases (created in the {\em enrollment phase})~\cite{biostar, china_hack} or from proposed
vector-sharing schemes~\cite{fawkes,foggy}. Once the attacker reconstructs these feature vectors, their
reconstructions must be good enough to be identified in both the {\em
  query} and {\em matching} phases for the images to be deanonymized.

\begin{figure*}
    \centering
    \includegraphics[width=0.95\textwidth]{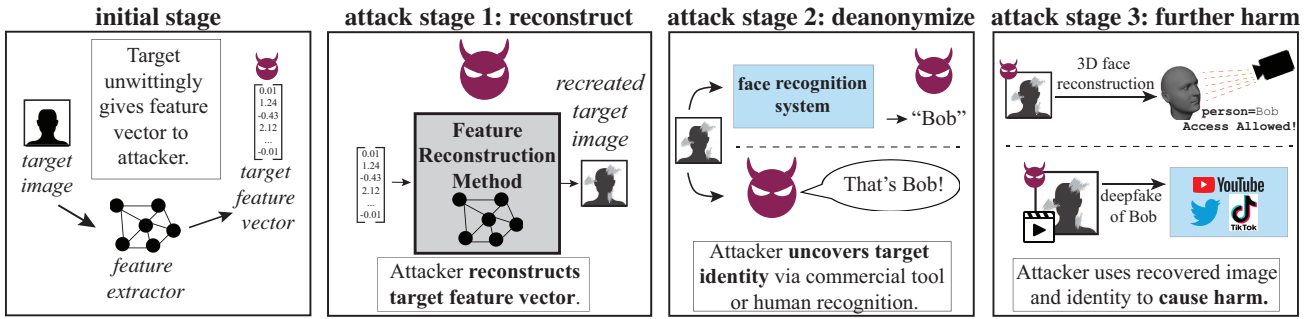}
    \caption{{\em The main phases of FVR-enabled attacks.}}
    \label{fig:cloak_sharing}
\end{figure*}

\section{Threat Model and Evaluation Metrics}\label{sec:metrics}

With this background in mind, we now formalize the threat model and
evaluation metrics for FVR-enabled deanonymization attacks on face images. We assume the
attacker attempts deanonymization via a facial recognition
engine, as manual reidentification of reconstructed images is
time-consuming and does not scale.  %

\subsection{Attack Stages and Assumptions}

We consider an adversary $\calA$, whose FVR-enabled deanonymization attack
operates in three stages, illustrated in
Figure~\ref{fig:cloak_sharing}. We assume $\calA$ has a feature vector
$v$ and blackbox access to the DNN $\calF$ that created it ({\em initial stage}). To conduct
the attack, $\calA$ inverts $v$ to recover $\tx$, an
approximation of real image $x$ of person $P$ ({\em stage 1});
deanonymizes $\tx$ to recover $P$'s identity ({\em stage 2});
and potentially conducts additional attacks using this knowledge
({\em stage 3}). Below, we describe the stages in more detail.

\para{Attack stage 1: reconstruction.} $\calA$ first
chooses a reconstructor $\calR_{\calF}$ and uses it to invert $\calF(x) = v$, producing $\tx=\calR_{\calF}(v)$. We
assume $\calA$ has a feature vector $v = \calF(x)$ generated from a face
  image $x$ from unknown person $P$ whose identity they wish to
  recover. Furthermore, we assume $\calA$ has black-box access to the feature extractor $\calF$ used to create
  $v$ and trains $\calR_{\calF}$ using dataset $\mathcal{X}_{\calR}, \mathcal{Y}_{\calR}$, $x \notin
  \mathcal{X}_{\calR}$. %

\para{Attack stage 2: deanonymization (primary attack).} Next, $\calA$
uses a {\em secondary facial recognition system $\calF'$} and/or {\em
  visual identification} to deanonymize $\tx$. In most cases, these
two methods are used together, but in limited cases $\calA$ could use visual identification alone --
if, for example, $P$ is a public figure who is easily recognized. In the more common scenario
that both $\calF'$ and visual identification are used, we assume
$\calF'$ is a commercial FR system that provides the top-$K$ potential identity
matches for an input image $\tx$ (see
Figure~\ref{fig:face_rec_overview}). $\cal{A}$ then chooses the best
visual match (if any) from the $K$ possibilities.

\para{Attack stage 3: integrity attack (secondary attack).} Though our
work measures the threat of deanonymization, $\cal{A}$ may conduct additional attacks after $\tx$ has been
deanonymized. For example, $\calA$ could use $\tx$ to generate
Deepfakes imitating $P$ or construct a 3D representation of $P$'s face to
fool biometric face recognition systems~\cite{3dmask_attack}. To conduct
each of these, the attacker needs only internet access and
a reasonably powerful computer. Code for both integrity attacks is available
freely online~\cite{deepfake_github, 3dmask_github}.

\subsection{Evaluation Metrics}
\label{subsec:metrics}

An efficient adversary would attempt deanonymization via a secondary
facial recognition engine $\calF' \ne \calF$ that produces a set of
potential identity matches for $\tx$. This attack technique naturally yields
two evaluation metrics, one based on $\calF'$'s {\bf top-$K$ matching accuracy}, and one based on the {\bf
  visual similarity}, for measuring the deanonymization risk posed by
different FVR methods. Both metrics are derived from key operational
components of modern facial recognition systems (see
Figure~\ref{fig:face_rec_overview}). Below, we describe the metrics in
detail and note their limited use in prior work.

\para{Metric 1: Top-$K$ Matching Accuracy.} The first phase of a
facial recognition system is {\em query matching}, in which the system produces
a set of $K$ possible identity matches for a queried image. For $\tx$
to be identified, the system must return the $P$'s true identity in
the top-$K$ set for $\tx$. Therefore, our first metric tests the frequency with which images
of $P$ appear in the top-$K$ sets produced for $\tx$ by secondary facial
recognition models, $\calF' \ne \calF$. We assume that $P$'s true
identity is enrolled in $\calF'$.
Prior FVR evaluations almost exclusively measure matching accuracy between $x$ and $\tx$ using the {\em
  inverted model $\calF$}, rather than a secondary model
$\calF'$. Top-$K$ match accuracy using $\calF$ provides a minimal
baseline for feature vector reconstruction performance. %

\para{Metric 2: Visual Similarity.} If images from correct class $P$ appear in the top-$K$ matches returned by $\calF'$, $\calA$
must identify them from the match set in the {\em matching}
phase of a facial recognition system. Thus, visual similarity between
$\tx$ and $P$'s true appearance is the second key
metric for assessing FVR-enabled deanonymization risk. In this
work, we employ a user study to measure this, which %
allows us to empirically assess whether a human could
deanonymize reconstructed images from a top-$K$ set. %
No prior FVR evaluation has used a user study to evaluate
reconstruction quality. Prior work primarily relies on SSIM~\cite{SSIM04}
to measure visual similarity. However, SSIM compares structural
similarity and fails to capture broader similarity between faces, especially if image backgrounds vary~\cite{zhang2018unreasonable}.

\section{Methodology}
\label{sec:method}
  \begin{figure}[t]
    \includegraphics[width=0.48\textwidth]{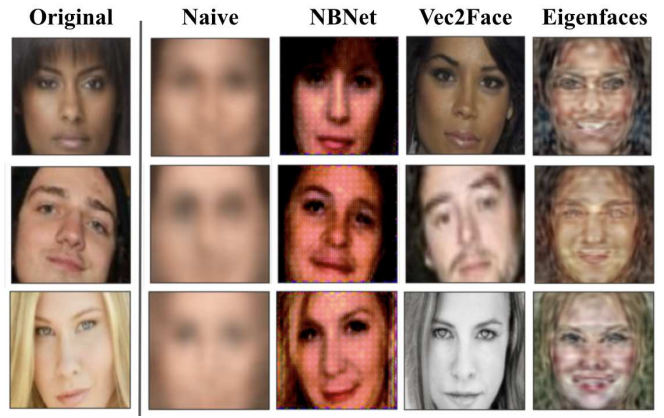}
    \caption{{\em Examples of images reconstructed by each method.}}
    \label{fig:all_reconstr}
  \end{figure}

Before presenting our evaluations using these metrics, we
describe the FVR methods, facial recognition models, and datasets used
in our analysis and provide a brief overview of the evaluation process.

\para{FVR Methods.} We measure the deanonymization success of four state-of-the-art FVR methods. These methods are chosen because they encompass all currently identified methods of FVR (i.e. parametric
and non-parametric), outperform older work, and have public
or partially public implementations. Below, we provide a brief
overview of each method. Examples of reconstructions from
each method are in Fig.~\ref{fig:all_reconstr}.

\texttt{NBNet}~\cite{nbnet} trains a model
  $\calR_{\calF}$ using a deconvolutional neural network architecture
  similar to DenseNet~\cite{dense}. We use the publicly available
  codebase~\cite{nbnetcode} and train $\calR_\calF$ using the {\em NBNet-B} architecture for $80$ epochs total with a batch size of
  $64$. We use pixel mean-absolute-error loss for 60 epochs and add perceptual loss for the last 20 epochs. 

  \texttt{Vec2Face}~\cite{Vec2Face} is generative adversarial net (GAN)-based
  and trains $\calR_{\calF}$ using a joint loss function balancing
  visual and feature space similarity between $x$ and $\tx$. The GAN
  architecture is based on PoGAN~\cite{karras2017progressive}, so we
  use the official PoGAN codebase to implement Vec2Face (for which no code was provided). Due to unanswered author correspondence, we omit $\ell_{biject}$ from the loss.\footnote{Since Vec2Faces assigns a comparatively small
  weight (0.01) to the $\ell_{biject}$ loss term, this omission does
  not significantly affect performance} We train using the recommended
  batch size scaling for PoGAN %
  and 15,000,000 training images.

 \texttt{Naive}~\cite{deeppoisoning21} is a convolutional neural net (CNN) built from stacked, inverted ResNet blocks~\cite{he2016deep}. We re-create the architecture based on the authors' detailed instructions and train each $\calR_\calF$ for $5$ epochs with a batch
  size of $128$. %

 \texttt{Eigenfaces}~\cite{eigenfaces} is a {\em nonparametric} FVR
 method that works by iteratively adding Gaussian
 blobs to $\tx$ which minimizes the feature space distance between the
 $\calF(\tx)$ and $v$. At each step, $\tx$ is normalized using
 pre-computed eigenface representations to encourage visual similarity
 between $x$ and $\tx$. We use author-provided code and run each
 $\calR_\calF$ procedure for $2000$ iterations with batch size $64$. %

\para{Facial Recognition Systems.} We use several facial recognition
(FR) models in our analysis. One set of local FR systems is used to
train parametric FVR methods and generate feature vectors for testing
($\calF$). The second set of commercial FR systems are used to evaluate Metric 1, the
top-$K$ matching accuracy of FVR methods in secondary models ($\calF'$).

{\bf {\em $\calF$: Models for FVR Training/Testing.}} We use two local FR models
to train and test FVR methods. Both are trained on the
MS1M~\cite{msceleb} dataset using the ArcFace~\cite{deng2019arcface}
loss function. One uses a ResNet50 backbone~\cite{he2016deep}, and the
other uses an EfficientNet backbone~\cite{tan2019efficientnet}. We
refer to these extractors as Res50 and Efficient respectively.
 
{\bf {\em $\calF'$: Models for Attack Evaluation.}} We measure
FVR-enabled deanonymization success using both local FR models
(i.e. Res50 and Efficient, described above) and
real-world FR systems, Microsoft Azure and Amazon Rekognition. For all systems, we
enroll a large set of labeled images to form
our ``matching set''. When unlabeled images are submitted for identification, the
systems return any potential matches found, along with a confidence
score for each.  %

\para{Datasets.} We use five well-known facial recognition datasets in our evaluation: one
to train models Res50 and Efficient, three
to train the parametric FVR methods $\calR$ (i.e. \texttt{NBNet},
\texttt{Vec2Face}, and \texttt{Naive}), and one to test reconstruction
performance. The datasets, and their uses, are in Table~\ref{tab:datasets}.

\begin{table}[ht]
    \centering
    \resizebox{0.48\textwidth}{!}{
      \begin{tabular}{c|c|r|r|l}
        \toprule
        \textbf{Use} & \textbf{Dataset} &
                                              \multicolumn{1}{c|}{\textbf{\textbf{\# Labels}}} & \multicolumn{1}{c|}{\textbf{\textbf{\# Images}}} & \multicolumn{1}{c}{\textbf{Citation}} \\ \midrule
        $\calF$ training & MS1M & 85742 & 1,776,809 &
                                                               ~\cite{msceleb}\\ \midrule
        \multirow{3}{*}{\begin{tabular}[c]{@{}c@{}} $\calR$ training \\and testing\end{tabular}} & FaceScrub & 530 & 57,838 & ~\cite{facescrub} \\
                         & VGGFace2 & 8,631 & 1,047,297 & ~\cite{vggface2}\\
                         & WebFace & 10,575 & 475,137 & ~\cite{webface} \\ \midrule
        $\calR$ testing & LFW & 5749 & 13233 & ~\cite{LFWTech} \\ \bottomrule
      \end{tabular}
    }
    \captionof{table}{{\em Datasets used for training and testing models $\calF$ and reconstructors $\calR$.}}
    \label{tab:datasets}
    \vspace{-3mm}
  \end{table}

\para{Evaluation Overview.} To perform our evaluation, we use the
local FR systems Res50 and Efficient as $\calF$ to train 18 FVR models $\calR_\calF$\footnote{We use 3 reconstruction methods (\texttt{NBNet},  \texttt{Vec2Face}, \texttt{Naive}) and 3 training datasets
   (WebFace, VGGFace2, FaceScrub) to invert each $\calF$,
   resulting in 9 $\calR_{\calF}$ for a single $\calF$ and 18 total
   $\calR_{\calF}$ across both $\calF$. Recall that Eigenfaces does
   not require training a $\calR$.}. We then reconstruct images using
 the trained FVR models (i.e. parametric methods) as well as the
 non-parametric \texttt{Eigenfaces}  method and measure
 deanonymization success on the reconstructions using Metrics 1 and 2 (\S\ref{subsec:metrics}). Our evaluation is structured as follows:

 \begin{packed_itemize}
 \item \S \ref{sec:feature_results} reports results on Metric 1, top-$K$ matching accuracy, using both local and
   commercial facial recognition systems. 
 \item \S \ref{sec:visual_results} reports results on Metric 2, visual
   similarity, via a survey-based user study.
 \item \S \ref{sec:discussion} synthesizes results and discusses the
   real-world threat of face image FVR, while proposing future work.
 \end{packed_itemize}

We report the results of reconstruction methods on Res50 feature
 vectors generated from FaceScrub images. Parametric FVR methods are
 trained using the WebFace dataset. Reconstruction results on
 Efficient feature vectors and other datasets are
 comparable and omitted for brevity. %

\begin{figure*}[t]
\centering
\begin{minipage}{.48\textwidth}
  \centering
  \includegraphics[width=.99\linewidth]{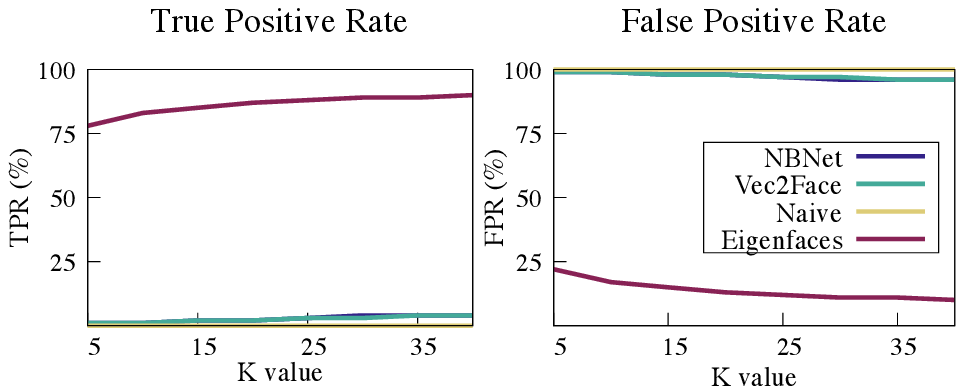}
  \captionof{figure}{{\em $tpr$ and $fpr$ rates for local $\calF$ facial
    recognition models when $N=1000$ and $K$ varies from $5$ to $40$.}}
  \label{fig:local_tpr1}
\end{minipage}%
\hspace{0.2cm}
\begin{minipage}{.48\textwidth}
  \centering
  \includegraphics[width=.99\linewidth]{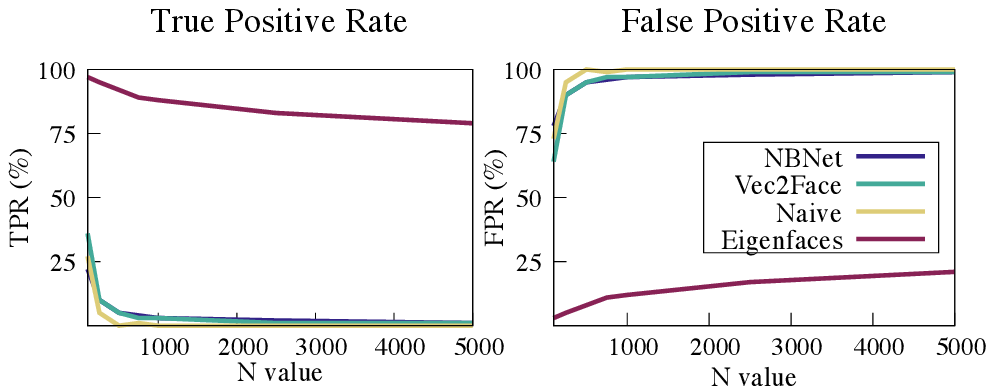}
  \captionof{figure}{{\em $tpr$ and $fpr$ rates for local $\calF$ facial
    recognition models when $K=25$ and $N$ varies from $50$ to $5000$.}}
  \label{fig:local_tpr2}
\end{minipage}
\end{figure*}

\begin{figure}[h]
  \includegraphics[width=0.5\textwidth]{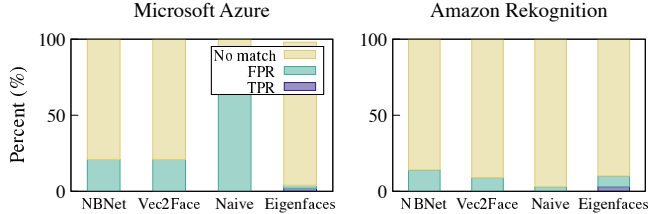}
  \captionof{figure}{{\em True positive rates, false positive rates, and
      ``no match'' rates for Azure and Rekognition.}}
  \label{fig:azure_aws1}
  \vspace{-5mm}
\end{figure}

\section{Evaluating Metric 1:  Top-$K$ Matching}
\label{sec:feature_results}

After $\calA$ uses FVR to recover a face image $\tx$, their first step
towards deanonymization is to run $\tx$ through a secondary facial recognition system $\calF'$. $\calF'$ returns the top-$K$
matches for a test image, the $K$ images with the highest similarity scores to $\tx$. Assuming $P$'s true identity is
enrolled in this system\footnote{A not-unreasonable assumption given
  the scale of modern facial recognition systems,
  c.f. Clearview.ai~\cite{hill_clearview}}, real images of $P$ should
appear in the top-$K$ options if the FVR attack is successful. 

\para{Measuring Metric 1.} We report the top-$K$ matching accuracy for
a FVR method using two measurements: the {\em true positive
  rate} ($tpr$: the proportion of $\tx$ for which $\calF'$ returned $>=1$ real
image of $P$ in top-$K$) and the {\em false positive rate} ($fpr$:
proportion of $\tx$ for which $\calF'$ returned only incorrect matches in the top-$K$). We first compute these on our local
Res50 and Efficient models, which allow more fine-grained
measurements, before running experiments on Azure and Rekognition.%

\subsection{Evaluation on Local Models}

To measure $tpr$ and $fpr$, we run experiments varying $K$ (match set size) and $N$ (number of enrolled
identities). This allows us to simulate real-world
facial recognition systems with different parameters, providing deeper insight.

\para{Results.} The nonparametric method (\texttt{Eigenfaces}) vastly outperforms parametric methods on both $tpr$ and
  $fpr$ in top-$K$ matching, regardless of $K$ or $N$ size.   Increasing $K$ and decreasing $N$ slightly improves
top-$K$ $tpr$ for all methods, but \texttt{Eigenfaces} still
dominates. As Figures~\ref{fig:local_tpr1} and~\ref{fig:local_tpr2}
show, \texttt{Eigenfaces} has a $tpr$ of at least $80\%$ across all $K$
and $N$ values, while the next best method, \texttt{Vec2Face} has a
$tpr$ of $40\%$ but only when the number of enrolled identities is
$N<50$. In a commercial system, $N$ would be much larger, so this
result overinflates \texttt{Vec2Face}'s performance.

Furthermore, \texttt{Vec2Face, NBNet}, and \texttt{Naive} have a high $fpr$ ($>90\%$) across all
values of $N$ and $K$. This means that deanonymization of images
reconstructed with these methods is difficult.

\subsection{Evaluation on Commercial Models}

For our tests on Rekognition and Azure, we
first enroll $N = 5000$ identities in each system by uploading
labeled face images to their databases. $500$ identities are drawn from the FaceScrub dataset, while
$4500$ are drawn from VGGFace2. The identities associated
with all reconstructions $\tx$ we test are also enrolled in the
system. %
To evaluate success, we query the reconstructions $\tx$ against the enrolled
identity database. We set $K=5$ (max), but since both systems only return
a match when the similarity score exceeds an unspecified threshold, some $\tx$
receive no matches. We ensure the small set of overlapping
classes between FaceScrub and VGGFace2 does not affect
our results. %

\para{Results.} %
As before, \texttt{Eigenfaces}
outperforms all other FVR methods on both Azure and
Rekognition (see Figure~\ref{fig:azure_aws1}). However, for
both systems, even \texttt{Eigenfaces} has $tpr <5\%$. The other methods fare
slightly better on Azure, but all have $tpr \le 1\%$. Both Azure and Rekognition have high $fpr$ for all methods
(Fig.~\ref{fig:azure_aws1}) -- up to $85\%$. Interestingly, many inversions from the same method are matched to the
  same false positive class. Table~\ref{tab:fp} shows the percent of non-unique false positives (i.e. same class) for each system/method
pair. Rekognition produces a higher percentage of non-unique
false positives than Azure on average. %

\begin{table}[h]
  \centering
  \resizebox{0.48\textwidth}{!}{
    \begin{tabular}{c|cccc} 
      \toprule
      \textbf{System} &  \textbf{NBNet} & \textbf{Vec2Face} & \textbf{Naive} & \textbf{Eigenfaces} \\ 
      \midrule
        \textbf{Rekognition} & 67\% & 54\% & 86\% &  77\% \\ 
      \midrule
      \textbf{Azure} & 24\% & 9\% & 99\% & 23\% \\
      \bottomrule
    \end{tabular}
  }
  \captionof{table}{{\em Percent of false positives which are from the same class for
      each method and system.}}
  \label{tab:fp}
  \vspace{-5mm}
\end{table}

\subsection{Discussion of Results}

Three out of four FVR methods evaluated have nontrival top-$K$
matching success, indicating a strong potential for full
deanonymization. In addition to this key result, we observe the
following notable behaviors.

\para{Nonparametric method performs best.} Images reconstructed using
\texttt{Eigenfaces} are matched at higher rates than images
reconstructed by other methods. This may be due to generalization errors in the parametric
reconstruction models, a well-known problem for machine learning models~\cite{demontis2019adversarial}. %
As nonparametric methods do not rely on a trained model, they do not suffer from this problem. 

\para{Noise from reconstruction process affects performance.} The many non-unique false positives observed in the commercial
systems likely indicates that the reconstruction models add
nontrivial, similar noise to all $\tx$. This noise causes the
consistent false positive classifications observed. Even
\texttt{Eigenfaces} has a $77\%$ ($23\%$) same-class false
positive rate in Rekognition (Azure). This presents an obstacle for
FVR-enabled deanonymization attacks. If the noise from the reconstruction
process is too strong, deanonymization will become more
difficult. However, future improvents to FVR techniques may mitigate
this problem.  %

\section{Evaluation of Metric 2: Visual Similarity}
\label{sec:visual_results}

If real images of $P$ appear in the top-$K$ set for $\tx$
(i.e. $tpr > 0$), $\calA$ must be able to identify them to complete
the deanonymization process
(c.f. Fig~\ref{fig:face_rec_overview}). Consequently, visual
similarity between $\tx$ and $P$ is a key component of successful
deanonymization attacks. In this section, we evaluate human ability to
match $\tx$ to the true target identity in top-$K$ matching sets for all four FVR methods using an
IRB-approved user study (IRB information omitted for anonymous
submission).

\para{Study Procedure.} The study asks participants to match
reconstructions to real images and then rank their confidence in the
match. In each question, participants are given the top $K$=5 matches
for $\tx$ generated using one of the four FVR methods. These are the
same top-$K$ matches produced by local models $\calF'$ used in the
prior section.  Participants indicate which (if any) of the top-$K$ images match
the identity of the person in $\tx$ and how confident they are in
their choice(s). Confidence options are given on a 5-point Likert scale, ranging from ``not
confident at all'' to ``very confident.'' Participants perform top-$K$
matching on $5$ images from each FVR method (20 images total). Random attention checks are included in the survey.

\begin{figure}[t]
  \centering
  \includegraphics[width=.8\linewidth]{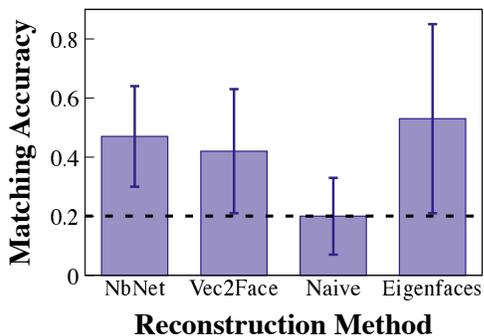}
  \captionof{figure}{{\em Proportion of survey responses with $\geq 1$
      match per method. The black dashed line represents the
      probability of a random correct guess.}} %
  \label{fig:survey2}
  \vspace{-4mm}
\end{figure}

\para{Study Participants.} We recruit 200 study participants via Prolific (\url{https://www.prolific.co/}).
Of our participants, $57\%$ identified as female ($39\%$ male; $4\%$ undisclosed).
The participants are all $18+$ years old spanning multiple age groups: 18-29 ($58\%$),
30-39 ($25\%$), 40-49 ($14\%$), 50-59 ($2\%$), 60+($1\%$).
The survey was designed to take 10 minutes on average,
and participants received $\$2$ as compensation. %

\para{Study Results.} Overall, users have the highest proportion of
successful matches (and thus deanonymizations) on \texttt{Eigenfaces} reconstructions
(Fig.~\ref{fig:survey2}). Since $K=5$, baseline random guessing
  results in $20\%$ matching accuracy, but participants successfully identify $50\%$ of matches on average.
The upper quartile for the matching accuracy of Eigenfaces is $85\%$
-- \emph{four times greater than baseline random guessing}. This
  clearly demonstrates that manual re-identification of the
  reconstructed images poses a very real threat -- a threat that can
  only increase as more robust FVR methods are developed.

Users are less confident in their responses to \texttt{Eigenfaces} questions (Table~\ref{tab:survey1}) than they are
for other successful methods. Respondents have the most confidence in their answers with \texttt{NBNet} and
\texttt{Vec2Face}. More than $90\%$ of respondents were confident to some degree with these two methods, and despite the poorer performance of the naive method, more than $70\%$ of respondents still felt confident about their selection. We conjecture that users feel more confident about their choices for \texttt{NBNet} and
\texttt{Vec2Face} due to the high visual quality of the images produced by these methods (see Figure~\ref{fig:all_reconstr}).

\section{Discussion}
\label{sec:discussion}

\para{Key Takeaways.} Our results demonstrate that FVR-enabled deanonymization attacks
  pose a real-world threat. Images reconstructed by the FVR methods we
  test produce successful top-$K$ matches in secondary
  facial recognition engines $\calF'$, and humans can successfully
  identify the target identity from these top-$K$ sets.

  Beyond this main finding, there are two other takeaways. First, nonparametric reconstruction methods are more flexible and, in our
  experience, suffer less generalization error than parametric methods. While visual image
  quality may suffer in the nonparametric setting (e.g. compare
  \texttt{Eigenfaces} to \texttt{Vec2Face} in Fig.~\ref{fig:all_reconstr}), future
  improvements may mitigate this problem. %
  Second, evaluations of future FVR methods should focus more on
  feature space similarity and less on visual similarity to provide more meaningful performance metrics. Since
  top-$K$ matching precedes visual matching in a practical attack leveraging facial
  recognition engines, current visual-focused metrics fall short in measuring
  the real-world threat. %

\para{Future Work.} The success of our FVR-enabled
attacks highlights a few key avenues for future work.

{\em Better defenses against FVR methods.} The non-negligible
probability of deanonymization shown in our work nullifies the claims that feature
vectors can be considered ``secure" in any meaningful, cryptographic
sense. As such, defenses against these attacks should be
developed. State-of-the-art defenses against FVR specifically (not evaluated in this work) assume the
attacker uses a parametric reconstruction method~\cite{deeppoisoning21,
  deepobfuscator19, nopeek20}. However, we found that the
nonparametric FVR method had the highest deanonymization success rate,
so future FVR defense work ought to broaden
its scope.

{\em Legislation.} 
Work along two lines is needed. 
First, the implications of face feature vectors as {\em both} personal
{\em and} biometric data need to be understood. For example, the EU's
General Data Protection Regulation (GDPR) defines \emph{biometric
  data} as ``Personal data resulting from specific technical
processing ... which allow or confirm [the] unique identification"
(Article 4(14))~\cite{GDPR}.  Personal and biometric data is subject
to more stringent regulations, and -- if feature vectors were to be
classified as such -- practitioners would need to take appropriate
measures to mitigate privacy risks associated with plaintext feature
vectors. As a second line of work, new legislation governing data
privacy laws should include face feature vectors. There is increasing
advocacy for the co-design of legislation and computer
science~\cite{legalpriv, legalpriv2} and new regulations should
account for privacy risks associated with face feature vectors and
facial recognition software.

\para{Ethics.} Our user study was approved by our local IRB and was
designed to maximally protect participant privacy. Furthermore, we
conducted our FVR attacks using public face datasets designed for
academic research use.

\begin{table}[t]%
  \centering
  \resizebox{0.49\textwidth}{!}{%
    \begin{tabular}{l|cccc}\toprule
      & \multicolumn{1}{l}{\bf NBNet} & \multicolumn{1}{l}{\bf Vec2Face} &
                                                                   \multicolumn{1}{l}{\bf
                                                                           Naive}
      & \multicolumn{1}{l}{\bf Eigenfaces} \\ \cmidrule{2-5}
      \multicolumn{1}{c|}{\textbf{Confident}}     & 0.92    & 0.94    & 0.77   & 0.85   \\
      \multicolumn{1}{c|}{\textbf{Not Confident}} & 0.08    & 0.06    &
                                                                      0.23   & 0.15  \\ \bottomrule
    \end{tabular}%
   }
  \caption{\em Average confidence in survey results from
      respondents. Though respondents ranked their confidence on a
      Likert scale, we categorize responses as
      ``confident'' (e.g. a response indicating some
      confidence) or ``not confident'' for easier presentation.}
    \label{tab:survey1}
    \vspace{-5mm}
\end{table}%

\newpage 
\bibliographystyle{IEEEtran}
\bibliography{references}

\end{document}